\documentclass[letterpaper]{JHEP3}

\usepackage{amsmath,amsthm, amssymb}

\newcommand{\be}{\begin{equation}}
\newcommand{\ee}{\end{equation}}
\newcommand{\ba}{\begin{array}}
\newcommand{\ea}{\end{array}}

\newcommand{\bea}{\begin{eqnarray}}
\newcommand{\eea}{\end{eqnarray}}
\newcommand{\bma}{\begin{matrix}}
\newcommand{\ema}{\end{matrix}}
\newcommand{\bpm}{\begin{pmatrix}}
\newcommand{\epm}{\end{pmatrix}}
\newcommand{\nn}{\nonumber}
\newcommand{\half}{\frac{1}{2}}
\newcommand{\mc}{\mathcal}
\newcommand{\beq}{\stackrel{\p\mc{M}}{=}}

\newcommand{\p}{\partial}

\newcommand{\rr}{\prime}
\newcommand{\ov}{\overline}
\newcommand{\wh}{\widehat}
\newcommand{\wt}{\widetilde}

\newcommand{\psibar}{\ov \psi}

\newcommand{\eps}{\varepsilon}
\newcommand{\ep}{\epsilon}

\newcommand{\al}{\alpha}

\newcommand{\la}{\lambda}
\newcommand{\da}{\delta}
\newcommand{\om}{\omega}
\newcommand{\Ga}{\Gamma}
\newcommand{\ga}{\gamma}

\newcommand{\qrq}{\quad\Rightarrow\quad}
\newcommand{\qlrq}{\quad\Leftrightarrow\quad}

\newcommand{\epbar}{\ov\ep}

\title{Tensor calculus for supergravity\\ on a manifold with boundary}
\author{ Dmitry V.~Belyaev \\
Deutsches Elektronen-Synchrotron, DESY-Theory\\ 
Notkestrasse 85, 22603 Hamburg, Germany\\
E-mail: \email{dmitry.belyaev@desy.de}
}
\author{ Peter van Nieuwenhuizen\\
C.~N.~Yang Institute for Theoretical Physics, SUNY at Stony Brook \\
Stony Brook, NY 11794-3840, USA\\
E-mail: \email{vannieu@max2.physics.sunysb.edu}
}
\date{\today}

\preprint{DESY-07-208\\ YITP-SB-07-33}

\abstract{
Using the simple setting of 3D $N=1$ supergravity, we show how the
tensor calculus of supergravity can be extended to manifolds
with boundary. 
We present an extension of the standard $F$-density formula 
which yields supersymmetric bulk-plus-boundary actions.
To construct additional \emph{separately} supersymmetric boundary actions,
we decompose bulk supergravity and bulk matter multiplets
into co-dimension one submultiplets. 
As an illustration we obtain the supersymmetric extension 
of the York-Gibbons-Hawking extrinsic curvature boundary term.
We emphasize that our construction does not require
any boundary conditions on off-shell fields. 
This gives a significant improvement over
the existing orbifold supergravity tensor calculus.
}

\dedicated{Dedicated to Julius Wess (1934-2007)}

\begin{document}

\numberwithin{equation}{section}


\section{Introduction} 

Supersymmetry (susy) and supergravity (sugra) were first
formulated in the 1970's as field theories in $x$-space
(the $x$-space or component approach).
A tensor calculus for 4D $N=1$ rigid susy,
with Poincar\'e or conformal symmetries, was developed
by Julius Wess and Bruno Zumino in their pioneering work \cite{RTC}.
For local susy (sugra), a tensor calculus for 4D $N=1$ models
was obtained in \cite{LTC1,LTC2}. 
At the same time, the superspace approach of Salam and Strathdee \cite{SS}
was extended to supergravity by Wess and Zumino \cite{SSsg} 
and was shown to be equivalent to the $x$-space tensor calculus 
approach \cite{LTCss}. 
Both approaches have been used since, and each has its own
virtues.

In all these studies, boundary effects were mostly ignored and
various total derivatives were simply dropped.
Already in the $x$-space approach, one calls a Lagrangian 
supersymmetric if its susy variation is a total derivative. 
In superspace, manipulations with susy-covariant
derivatives $D_\al$ often produce total $x$-space derivatives which
are again discarded under the $x$-space integration. 
One cannot do so in the presence of boundaries in $x$-space, which
is why the superspace and tensor calculus approaches are not
obviously extendable to a manifold with boundary.

Susy models in the presence of $x$-space boundaries have been studied
before. Boundary terms for open fermionic strings \cite{pdv} and 
the Casimir effect in 4D susy theories \cite{igarashi} 
were among the first considered. 
(For a flavor of other models discussed over the years, see
\cite{bmods}.)
Already in \cite{pdv} it was argued that one needs boundary
conditions (BC) to maintain (at least part of) susy in the presence
of a boundary, and that the BC must, in turn, be left invariant under 
susy transformations 
(that is, form a ``susy orbit'' \cite{LRN}).
This approach, which we will call ``susy with BC,''
was used in most works on susy in the presence of 
boundaries.

In a recent analysis of \cite{LRN,VV}, the BC required by 
the Euler-Lagrange variational principle,
were considered together with the BC needed to maintain
susy of the actions. The orbit of all BC was constructed,
and the functional space of off-shell fields was
defined by the set of all constraints. 
Here we take a completely opposite point of view: 
we develop an approach to rigid and local susy in which off-shell
fields are totally unconstrained.

Our approach gives classical\footnote{
At the quantum level, local susy is replaced by BRST symmetry,
but the same approach can be followed \cite{LRN,VV}.
} 
bulk-plus-boundary actions that are susy (under
a half of bulk susy) without using any BC on fields.
We call our approach ``susy without BC'' to contrast it with the
``susy with BC'' approach used so far.\footnote{
We will impose BC on \emph{symmetry parameters}, but not on fields.
Of course, BC on fields follow upon applying the variational principle
to our actions, but these BC are not needed in the proof of susy
of the actions. Whether these BC form susy orbits~\cite{LRN,VV} 
is a separate issue that we will discuss elsewhere \cite{tap}.
}
For rigid susy, the validity of this approach has already been
established by one of us in \cite{db1}. 
The key ingredient used there, which made the construction
particularly simple, was the co-dimension one decomposition of 
(rigid) superfields \cite{rigidDd}.
In this article, we will give a first complete realization of
this approach in the case of \emph{local} susy (sugra).
We restrict our discussion to a 3D space-time and 
show how the complete tensor calculus for 3D $N=1$ local susy 
can be extended to take boundaries into account. 
Co-dimension one decomposition of the bulk susy multiplets will
play an essential role in our construction.
An extension of our construction to higher
dimensions and its superspace realization will be discussed
elsewhere \cite{tap}.


Understanding supergravity on a manifold with boundary is an
interesting mathematical problem. It is also important
for various physical models that have appeared in the past decade.
Notably, the 11D Horava-Witten (HW) construction \cite{hw} and the 5D
Randall-Sundrum (RS) scenario \cite{rs} 
(whose minimal supersymmetrization
was achieved in \cite{susyRS}).\footnote{
The HW and (susy) RS models are usually discussed 
in the ``upstairs picture''
(on the $S^1/\mathbb{Z}_2$ orbifold). The alternative ``downstairs
picture'' (on a manifold with boundary) approach to these models
was considered, for example,
in \cite{moss} and \cite{db2}, respectively. 
Here we adhere to the ``downstairs picture'' description.
}
In these models, one starts from a (standard) bulk supergravity action 
and tries to construct a boundary action (involving, in general,
additional boundary-localized fields) that makes the whole system
supersymmetric (under a half of bulk susy, with the other
half being spontaneously broken by the presence of the boundary).
As of now, most approaches to constructing such
susy bulk-plus-boundary actions have relied on certain
approximations. For example, 
\begin{enumerate}
\item
the 11D HW action is susy
only to a certain order in the expansion parameter $\kappa^{2/3}$ 
\cite{hw,moss};
\item
the 5D orbifold supergravity tensor calculus of \cite{zucker,kugo} 
relies on using standard orbifold ``odd=0'' BC
which, in general, are incompatible with the BC
one derives from the variational principle \cite{db4};
\item
the 5D constructions of \cite{bbf}, which incorporate BC
following from the variational principle,
are worked out only to lowest fermi
order.
\end{enumerate}
We hope that our approach, which works without any approximations
or assumptions, will help to bring these constructions to completion.

We base our construction on the existing tensor calculus for
3D $N=1$ and 2D $N=(1,0)$ supergravity. This tensor calculus
was worked out by Uematsu \cite{ue3,ue2}, following the
4D $N=1$ results of \cite{LTC1}.
In these derivations, conformal sugra plays a fundamental
role, but we consider only Poincar\'e sugra in this article.

Our construction will consist of the following steps.

First, we analyze the algebra of supergravity gauge transformations.
We recall why, in the presence of a boundary, one can (typically)
preserve only half of bulk susy, and prove that the restriction
to this half of susy reduces the whole 3D $N=1$ gauge algebra
to the standard 2D $N=(1,0)$ gauge algebra, without imposing any
BC on fields.
We note that the analysis becomes particularly simple in a special 
Lorentz gauge (which is opposite to the standard Kaluza-Klein choice) 
and we adopt that gauge from then on.
As a consequence, the preserved half of susy transformations 
gets modified by a compensating Lorentz transformation.

Second, we perform a co-dimension one decomposition of the
3D supergravity tensor calculus. This gives, in particular,
the induced supergravity multiplet that is necessary for 
constructing \emph{separately susy} 
boundary actions. The decomposition does not rely on using any
BC (like ``odd=0'' BC used in \cite{zucker,kugo})
and is applicable to any hypersurface parallel to the boundary.

Third, we show that on a manifold with boundary, the standard 3D 
$F$-density formula must be extended by the addition of a boundary 
$A$-term. The extended $F$-density formula automatically gives
bulk-plus-boundary actions that are susy (under the half
of bulk susy) without using any BC on fields.
We also write the extended $F$-density in terms
of the co-dimension one submultiplets.

To illustrate the construction, we finally apply the extended $F$-density
formula to the 3D $N=1$ scalar curvature multiplet. This will show that
the minimal susy bulk-plus-boundary action, with the standard
3D $N=1$ supergravity in the bulk, does not include the York-Gibbons-Hawking 
term \cite{ygh}. The latter comes as a part of a separately susy
boundary action that one needs to add in order to relax field
equations which would otherwise be too strong.

\section{Co-dimension one gauge algebra}

In this section, we will show how the 3D $N=1$ supergravity gauge
algebra\footnote{
The gauge algebra of 4D $N=1$ sugra was first discussed in
\cite{FPVN}, and its closure if auxiliary fields are included 
was discussed in \cite{LTC1,LTC2}.
}
reduces naturally to the 2D $N=(1,0)$ supergravity gauge
algebra on the boundary, as well as on 
co-dimension one slices parallel to the boundary.

\subsection{3D $N=1$ gauge algebra}

The gauge transformations of the 3D $N=1$ (off-shell) 
Poincar\'e supergravity
are the Einstein (general coordinate) transformation $\da_E(\xi^M)$,
the local Lorentz transformation $\da_L(\la^{AB})$ and the
susy transformation $\da_Q(\ep)$. The complete gauge
algebra reads\footnote{
Our conventions are: $M$, $N$ are curved 3D indices, 
$A$, $B$ are flat 3D indices, with decomposition
$M=(m,3)$ and $A=(a,\hat 3)$. The 3D gamma matrices satisfy
$\ga^A\ga^B=\ga^{AB}+\eta^{AB}$
with $\eta^{AB}=(-++)$ and $\ga^A\ga^B\ga^C=\ga^{ABC}+
\eta^{AB}\ga^C+\eta^{BC}\ga^A-\eta^{AC}\ga^B$
with $\ga^{ABC}=\eps^{ABC}$. Our spinors are Majorana;
$\psibar=\psi^{\rm T}C$, $C^{\rm T}=-C$, 
$C\ga^A C^{-1}=-(\ga^A)^{\rm T}$.
Einstein transformations yield
$\da_\xi e_M{}^A=\xi^N\p_N e_M{}^A+e_N{}^A\p_M\xi^N$, etc.;
Lorentz and susy transformations are given in (\ref{Lorentz}),
(\ref{susy1}) and (\ref{susy2}).
}
\bea
\label{3Dga}
& [\da_{E}(\xi_1^M)+\da_L(\la_1^{AB})+\da_Q(\ep_1), \;\;
\da_{E}(\xi_2^M)+\da_L(\la_2^{AB})+\da_Q(\ep_2)] & \nn\\[5pt]
&= \da_{E}(\xi_\text{comp}^M)+\da_L(\la_\text{comp}^{AB})
+\da_Q(\ep_\text{comp}) &
\eea
where the composite parameters are 
\bea
\label{comp}
\xi_\text{comp}^M &=& 2(\ov\ep_2\ga^M\ep_1)+\Big[
\xi_2^N\p_N\xi_1^M-(1\leftrightarrow2) \Big] \nn\\[5pt]
\la_\text{comp}^{AB} &=& 2(\ov\ep_2\ga^N\ep_1)\wh\om_N{}^{AB}
+(\ov\ep_2\ga^{AB}\ep_1)S +\Big[
\xi_2^N\p_N\la_1^{AB}
+\la_2^A{}_C\la_1^{CB}
-(1\leftrightarrow2) \Big] \nn\\
\ep_\text{comp} &=& -(\ov\ep_2\ga^M\ep_1)\psi_M+\Big[
\xi_2^N\p_N\ep_1
+\frac{1}{4}\la_2^{AB}\ga_{AB}\ep_1
-(1\leftrightarrow2) \Big]
\eea
with $\ga^M=\ga^A e_A{}^M$. 
The composite parameters depend explicitly 
on the fields of the 3D supergravity multiplet $(e_M{}^A, \psi_M, S)$,
with $e_A{}^M$ being the inverse of $e_M{}^A$ and $\wh\om_{MAB}$
being the supercovariant spin connection (see (\ref{spincon})). The algebra is realized on 
the supergravity multiplet itself, as well as on other 3D multiplets
such as the 3D scalar multiplet $\Phi_3(A)=(A,\chi,F)$.

\subsection{Einstein boundary condition}

We are interested in constructing supersymmetric bulk-plus-boundary
actions of the form
\bea
S=\int_{\mc{M}} d^3x \mc{L}_3
+\int_{\p\mc{M}} d^2x \mc{L}_2
\eea
For notational simplicity,\footnote{
Our choice of coordinates $x^M$ does not impose an Einstein
gauge as it does not restrict $\xi^M(x)$. It also does not
imply that our boundary has to be flat, because it places no 
restrictions on (intrinsic or extrinsic) curvature.
}
we choose the coordinates $x^M$ 
in such a way that the boundary $\p\mc{M}$ is at $x^3=0$ and that $x^3>0$
in the bulk $\mc{M}$. 
The boundary has coordinates $x^m=(x^0,x^1)$.
Under Einstein transformations, $\mc{L}_3$ is assumed to be a density,
$\da_\xi\mc{L}_3=\p_M(\xi^M\mc{L}_3)$, so that
\bea
\da_\xi S=\int_{\p\mc{M}} d^2x \Big(
-\xi^3\mc{L}_3+\da_\xi\mc{L}_2 \Big)
\eea
The standard way to achieve $\da_\xi S=0$ is
to impose a BC on the Einstein parameter,
\bea
\label{xi3}
\xi^3 \beq 0
\eea
and take $\mc{L}_2$ to be a density under the induced Einstein 
transformations, $\da_\xi\mc{L}_2=\p_m(\xi^m\mc{L}_2)$.
(We assume that the total $\p_m$ derivative integrates to zero on 
the boundary.) In principle, one could investigate other ways
to achieve $\da_\xi S=0$ without imposing the BC (\ref{xi3}).
In this article, however, we will assume that this BC on the 
parameter $\xi^M$ has to be imposed.

\subsection{The unbroken half of bulk susy}

Consistency of the gauge algebra (\ref{3Dga}) with the BC
(\ref{xi3}) requires \cite{VV}
\bea
\label{e120}
\xi^3_\text{comp}\beq 0 \qlrq
(\epbar_2\ga^A\ep_1)e_A{}^3 \beq 0
\eea
It is convenient to choose a special Lorentz gauge,\footnote{
Note that the gauge $e_a{}^3=e_m{}^{\hat 3}=0$ is opposite 
to the standard Kaluza-Klein choice \cite{KK}, $e_{\hat 3}{}^m=e_3{}^a=0$.
It is the analog of the ``time gauge'' introduced by 
Schwinger~\cite{time1}
for the Hamiltonian analysis of gravity. (For the Hamiltonian 
analysis of the Dirac action in a curved space it was used by 
Kibble~\cite{time2},
and for the Hamiltonian formulation of 4D $N=1$ supergravity 
it was used in \cite{hsg1}). 
In more mathematical terms, this gauge corresponds to the choice 
of a \emph{surface-compatible frame} \cite{isenberg}.
Its usefulness in the setting of supergravity
on a manifold with boundary was emphasized in \cite{db2}.
}
\bea
\label{gauge}
\boxed{ \rule[-5pt]{0pt}{16pt} \quad
e_a{}^3=0 \qrq e_m{}^{\hat 3}=0
\quad}
\eea
both on $\p\mc{M}$ and in $\mc{M}$. (We shall later comment
on the case when one does not impose this gauge.)
As $e_{\hat 3}{}^3$ is 
non-zero, the BC (\ref{e120}) now reduces to a field-independent requirement
\bea
\epbar_2\ga^{\hat 3}\ep_1 \beq 0
\eea
Introducing projectors $P_{\pm}=\half(1\pm\ga^{\hat 3})$
and defining $\ep_{\pm}=P_{\pm}\ep$, we solve this BC
by imposing (without loss of generality) the following BC
on the susy parameter $\ep$,
\bea
\label{susybc}
\ep_{-}\beq 0 \qlrq \ep\beq\ep_{+}
\eea
The half of susy that is not broken by the boundary
satisfies
\bea
\ep_{+}=P_{+}\ep_{+}, \quad
\epbar_{+}=\epbar_{+}P_{-}, \quad
\ga^{\hat 3}\ep_{+}=\ep_{+}, \quad
\epbar_{+}=-\epbar_{+}\ga^{\hat 3}
\eea
The other half, parametrized by $\ep_{-}$, is broken by the boundary. 
It could, in principle, be restored by introducing appropriate 
Goldstone fields on the boundary, which would show that the breaking 
is spontaneous. However, in this article, 
we will only be interested in preserving the $\ep_{+}$ susy.

\subsection{Modified $\ep_{+}$ susy}

The gauge condition (\ref{gauge}) is invariant under arbitrary
$\xi^m$ and $\la^{ab}$ transformations, but not under 
$\la^{a\hat 3}$ and $\ep_{+}$ ones. 
Only a particular combination of $\la^{a\hat 3}$ and $\ep_{+}$
transformations survives in this gauge. 
We, therefore, introduce a \emph{modified} $\ep_{+}$ susy 
transformation,
\bea
\label{msusy}
\boxed{ \rule[-5pt]{0pt}{16pt} \quad
\da_Q^\rr(\ep_{+})=\da_Q(\ep_{+})
+\da_L(\la_{a\hat 3}^\rr=-\epbar_{+}\psi_{a-})
\quad}
\eea
which satisfies $\da_Q^\rr(\ep_{+})e_m{}^{\hat 3}=0$.
(We will use the notation $\da^\rr_\ep\equiv\da_Q^\rr(\ep_{+})$.)
It is this $\ep_{+}$ susy transformation that we will
use in the following constructions.

\subsection{The reduced gauge algebra}

We claim that 
the surviving gauge transformations, $\da_E(\xi^m)$, $\da_L(\la^{ab})$,
and $\da_Q^\rr(\ep_{+})$, form a subalgebra of the 3D $N=1$ supergravity
gauge algebra that is isomorphic to the (standard) 2D $N=(1,0)$ supergravity
gauge algebra. The non-trivial part of the proof concerns
the commutator of two (modified) $\ep_{+}$ susy transformations.
We find
\bea
\label{Q12a}
[\da_Q^\rr(\ep_{1+}),\da_Q^\rr(\ep_{2+})]
=\da_E(\xi^M)+\da_L(\la^{AB})+\da_Q(\ep)+\da_L(\wt\la_{a\hat 3})
\eea
where \footnote{
The extra terms in $\la_{ab}$ and $\ep$ arise from the terms
$(\la^\rr_2)_a{}^{\hat 3}(\la^\rr_1)_{\hat 3b}$
and
$\half\la^\rr_2{}^{a\hat 3}\ga_{a\hat 3}\ep_{1+}$
in (\ref{comp}) upon
using the Fierz identities
$(\epbar_{+}\psi_{-})(\ov\phi_{-}\eta_{+})=
-\half(\epbar_{+}\ga^c\eta_{+})(\ov\phi_{-}\ga_c\psi_{-})$
and
$(\epbar_{+}\psi_{-})\phi_{-}
=-(\epbar_{+}\phi_{-})\psi_{-}$.
}
\bea
\xi^m &=& 2(\epbar_{2+}\ga^a\ep_{1+})e_a{}^m, \quad \xi^3=0, \quad
\ep = -\half\xi^n\psi_n+\half\xi^n\psi_{n-}
\nn\\[5pt]
\la_{ab} &=& \xi^n\Big[
\wh\om_{nab}-\half\psibar_{a-}\ga_n\psi_{b-} \Big], \quad
\la_{a\hat 3} = \xi^n\Big[
\wh\om_{na\hat 3}+\half S e_{na} \Big] 
\eea
The extra composite Lorentz transformation with
\bea
\wt\la_{a\hat 3}=-\epbar_{2+}\da_Q^\rr(\ep_{1+})\psi_{a-}
-(1\leftrightarrow 2)
\eea
arises because the compensating Lorentz transformation in (\ref{msusy})
is field-dependent. We see immediately that the (composite)
$\ep_{-}$ vanishes identically (without imposing $\psi_{m-}=0$),
thanks to the contribution from the compensating Lorentz transformation. 
Using the results of the next section, 
one finds that \cite{tap}
\bea
\la_{a\hat 3}+\wt\la_{a\hat 3}=\half\xi^n\psibar_{n+}\psi_{a-}, \quad
\wh\om_{nab}-\half\psibar_{a-}\ga_n\psi_{b-}=\wh\om_{nab}^{+}
\eea
where $\wh\om_{nab}^{+}$ is the standard supercovariant connection
constructed out of $e_m{}^a$ and $\psi_{m+}$. This brings (\ref{Q12a})
to the form
\bea
[\da_Q^\rr(\ep_{1+}),\da_Q^\rr(\ep_{2+})]
=\da_E(\xi^m)+\da_L(\la_{ab}=\xi^n\wh\om_{nab}^{+})
+\da_Q^\rr(\ep_{+}=-\half\xi^n\psi_{n+})
\eea
which is the standard form of the 2D $N=(1,0)$ (local) susy algebra.
We emphasize that we have identified this subalgebra without
imposing any boundary conditions on supergravity fields.
Accordingly, this identification works for any hypersurface 
$x^3=\text{const}$ parallel to the boundary $\p\mc{M}$.

\section{Co-dimension one submultiplets}

Having proved that the 3D $N=1$ supergravity gauge algebra
reduces to the 2D $N=(1,0)$ supergravity gauge algebra on the
hypersurfaces parallel to the boundary, we are guaranteed
that the 3D multiplets can be decomposed into a set of 2D
submultiplets. In this section, we will describe these submultiplets 
for the 3D supergravity and the 3D scalar multiplets.

\subsection{3D supergravity multiplet}

The 3D supergravity multiplet, $(e_M{}^A,\psi_M,S)$,
enjoys the following susy transformations,
\bea
\label{susy1}
\da_\ep e_M{}^A=\epbar\ga^A\psi_M, \quad
\da_\ep\psi_M=2\wh D_M\ep, \quad
\da_\ep S=\half\epbar\ga^{MN}\wh\psi_{MN}
\eea
where $\wh\psi_{MN}=\wh D_M\psi_N-\wh D_N\psi_M$ is the supercovariant 
gravitino field strength and
\bea
\wh D_M\ep = D_M(\wh\om)\ep+\frac{1}{4}\ga_M\ep S, \quad
\wh D_M\psi_N = D_M(\wh\om)\psi_N-\frac{1}{4}\ga_N\psi_M S
\eea
The covariant derivatives $D_M$ are only Lorentz covariant, so that
\bea
D_M(\wh\om)\psi_N=\p_M\psi_N+\frac{1}{4}\wh\om_{MAB}\ga^{AB}\psi_N
\eea
and the supercovariant spin connection is given by
\bea
\label{spincon}
\wh\om_{MAB}=\om(e)_{MAB}+\kappa_{MAB}, \quad
\kappa_{MAB}=\frac{1}{4}(\psibar_M\ga_A\psi_B-\psibar_M\ga_B\psi_A
+\psibar_A\ga_M\psi_B) \nn\\
\om(e)_{MAB}=\half(C_{MAB}-C_{MBA}-C_{ABM}), \quad
C_{MN}{}^A=\p_M e_N{}^A-\p_N e_M{}^A
\eea
where we use the standard conversion of indices, $\psi_A=e_A{}^M\psi_M$,
etc. The supercovariant spin connection has the following
susy transformation,
\bea
\da_\ep\wh\om_{MAB}=\half\epbar(\ga_B\wh\psi_{MA}-\ga_A\wh\psi_{MB}
-\ga_M\wh\psi_{AB})-\half(\epbar\ga_{AB}\psi_M)S
\eea
Under a 3D Lorentz transformation, we have
\bea
\label{Lorentz}
\da_\la e_M{}^A=\la^{AB}e_{MB}, \quad
\da_\la\psi_M=\frac{1}{4}\la^{AB}\ga_{AB}\psi_M, \quad
\da_\la S=0, \quad
\da_\la\wh\om_{MAB}=-D(\wh\om)_M\la_{AB}
\eea
These Lorentz transformations will play a role
as the (modified) $\ep_{+}$ susy transformation 
(\ref{msusy}) involves a compensating Lorentz transformation.

\subsection{Co-dimension one split}

To identify co-dimension one submultiplets of the supergravity
multiplet, we first split the indices, $M=(m,3)$, $A=(a,\hat 3)$,
and the spinors, $\ep=\ep_{+}+\ep_{-}$. The resulting component
fields (and parameters) can be formally assigned parities 
(in a way consistent with the susy transformations) as follows,
\bea
\label{parity}
\begin{array}{r@{\quad}
l@{\quad}l@{\quad}l@{\quad}l@{\quad}
l@{\quad}l@{\quad}l@{\quad}l@{\quad}l}
\text{even:} & e_m{}^a & e_3{}^{\hat 3} & 
\om_{mab} & \om_{3a\hat 3} &
{} & \psi_{m+} & \psi_{3-} & \ep_{+} & \p_m\\
\text{odd:}  & e_3{}^a & e_m{}^{\hat 3}=0 & 
\om_{3ab} & \om_{ma\hat 3} &
S & \psi_{m-} & \psi_{3+} & \ep_{-}=0 & \p_3
\end{array}
\eea
(The vanishing of $e_m{}^{\hat 3}$ and $\ep_{-}$ correspond to
our Lorentz gauge choice (\ref{gauge}) 
and the restriction (\ref{susybc}) on susy, respectively.)
Co-dimension one multiplets will have definite parities
as well.

In general, the induced metric on the $x^3=\text{const}$ slices
is $g_{mn}=e_m{}^a e_{na}+e_m{}^{\hat 3} e_{n\hat 3}$. With our
choice of the Lorentz gauge, however, we have $g_{mn}=e_m{}^a e_{na}$,
so that $e_m{}^a$ is the induced vielbein. One can also easily check
that $\om(e)_{mab}$ coincides with the torsion-free spin connection
constructed out of $e_m{}^a$, whereas $\om(e)_{ma\hat 3}e_n{}^a$
coincides, up to a convention-dependent sign, with the extrinsic
curvature tensor \cite{db2}. We fix the sign by defining\footnote{
The extrinsic curvature is usually defined by
$K_{MN}=\pm P_M{}^K P_N{}^L \nabla_K n_L$ where 
$P_M{}^K=\da_M{}^K-n_M n^K$ and 
$\nabla_K n_L=\p_K n_L-\Ga_{KL}{}^S n_S$.
In our gauge and with our choice of coordinates, 
$n_M=(0,0,-e_3{}^{\hat 3})$
and
$K_{mn}=\mp\Ga_{mn}{}^3 n_3=\pm\Ga_{mn}{}^3 e_3{}^{\hat 3}$.
The vielbein postulate yields $\Ga_{mn}{}^3 e_3{}^{\hat 3}
=-\om_{ma}{}^{\hat 3} e_n{}^a$.
(See appendices in \cite{VV} and \cite{db2} 
for more details and references.)
Our sign choice is then 
$K_{MN}=- P_M{}^K P_N{}^L \nabla_K n_L$.
}
\bea
K_{mn}=\om(e)_{ma\hat 3}e_n{}^a
\eea
In our gauge, $e_m{}^{\hat 3}=e_a{}^3=0$, we have 
$e_m{}^a e_a{}^n=\da_m{}^n$, $e_a{}^m e_m{}^b=\da_a{}^b$ and 
$e_3{}^{\hat 3} e_{\hat 3}{}^3=1$, as well as
\bea
&& \ga_m=e_m{}^a\ga_a, \quad 
\ga_3=e_3{}^a\ga_a+e_3{}^{\hat 3}\ga_{\hat 3}, \quad
\ga^m=\ga^a e_a{}^m+\ga^{\hat 3} e_{\hat 3}{}^m, \quad
\ga^3=\ga^{\hat3} e_{\hat 3}{}^3 \nn\\[5pt]
&& \psi_a=e_a{}^m\psi_m, \quad
\psi_{\hat 3}=e_{\hat 3}{}^m\psi_m+e_{\hat 3}{}^3\psi_3
\eea
We will also use $K_{ma}=\om(e)_{ma\hat 3}$ and $K_{ba}=e_b{}^m K_{ma}$.
Noting that $\wh\om_{ma\hat 3}$ is \emph{not} supercovariant under
the (modified) $\ep_{+}$ susy, we define the supercovariant extrinsic
curvature tensor as
\bea
\label{defK}
\wh K_{ma}=\wh\om_{ma\hat 3}-\half\psibar_{m+}\psi_{a-}
\eea
Using 
$\psibar_m\psi_a=\psibar_{m+}\psi_{a-}+\psibar_{m-}\psi_{a+}$
and 
$\psibar_m\ga_{\hat 3}\psi_a=-\psibar_{m+}\psi_{a-}+\psibar_{m-}\psi_{a+}$,
we find that
\bea
\label{scEC}
\wh K_{ma}=K_{ma}+\frac{1}{4}(
\psibar_m\ga_a\psi_{\hat 3}-\psibar_m\psi_a
+\psibar_a\ga_m\psi_{\hat 3})
\eea
As the bosonic extrinsic curvature tensor is symmetric, $K_{ab}=K_{ba}$, 
the supercovariant extrinsic curvature tensor
is symmetric as well, $\wh K_{ab}=\wh K_{ba}$. 

\subsection{Induced supergravity multiplet}

Under the (modified) $\ep_{+}$ susy (\ref{msusy}), 
the induced vielbein transforms as follows,
\bea
\da^\rr_\ep e_m{}^a=\epbar_{+}\ga^a\psi_{m+}
\eea
(The compensating Lorentz transformation does not contribute here
as $\la^\rr{}^{a\hat 3} e_{m\hat 3}$ vanishes in our gauge.)
The variation of $\psi_{m+}$ gives
\bea
\da^\rr_\ep\psi_{m+}=2(\p_m+\frac{1}{4}\wh\om_{mab}\ga^{ab})\ep_{+}
+\half\la^\rr_{a\hat 3}\ga^{a\hat 3}\psi_{m-}
\eea
where $\la^\rr_{a\hat 3}=-\epbar_{+}\psi_{a-}$. Performing
the following decomposition,
\bea
\wh\om_{mab} &=& \wh\om_{mab}^{+}+\kappa_{mab}^{-}, \quad
\kappa_{mab}^{-}=\frac{1}{4}(\psibar_{m-}\ga_a\psi_{b-}
-\psibar_{m-}\ga_b\psi_{a-}+\psibar_{a-}\ga_m\psi_{b-})
\nn\\
\wh\om_{mab}^{+} &=& \om(e)_{mab}+\kappa_{mab}^{+}, \quad
\kappa_{mab}^{+}=\frac{1}{4}(\psibar_{m+}\ga_a\psi_{b+}
-\psibar_{m+}\ga_b\psi_{a+}+\psibar_{a+}\ga_m\psi_{b+})
\eea
we observe that $\wh\om_{mab}^{+}$ is the (standard) supercovariant
spin connection for the 2D (induced) vielbein $e_m{}^a$.
Defining the 2D (Lorentz) covariant derivative as
\bea
D^\rr_m(\wh\om^{+})\ep=\p_m\ep+\frac{1}{4}\wh\om_{mab}^{+}\ga^{ab}\ep
\eea
we arrive at
\bea
\da^\rr_\ep\psi_{m+}=2 D^\rr_m(\wh\om^{+})\ep_{+}
+\half\kappa_{mab}^{-}\ga^{ab}\ep_{+}
+\half\la^\rr_{a\hat 3}\ga^{a\hat 3}\psi_{m-}
\eea
We claim that the last two terms cancel each other. 
To prove this, we first observe that the antisymmetrization in
any three 2D vector indices gives zero, $[abc]=0$, which yields
\bea
\kappa_{mab}^{-}=\frac{1}{2}\psibar_{a-}\ga_m\psi_{b-}
\eea
Second, the identity $\ga^{ab}=\ep^{ab\hat 3}\ga_{\hat 3}$ accounts for
a useful trick,
\bea
\ga^{ab}\ep_{+}(\psibar_{a-}\ga_m\psi_{b-})
=-\ep_{+}(\psibar_{a-}\ga_m\ga^{ab}\psi_{b-})
\eea
Finally, gamma-matrix algebra reduces the last term to $2\ep_{+}(\psibar_{a-}\ga^a\psi_{m-})$
and the Fierz transformation gives
\bea
\ga^{ab}\ep_{+}(\psibar_{a-}\ga_m\psi_{b-})
=-2\ga^a\psi_{m-}(\epbar_{+}\psi_{a-})
\eea
which proves our statement and gives us the final result,
\bea
\label{isg}
\boxed{ \rule[-5pt]{0pt}{16pt} \quad
\da^\rr_\ep e_m{}^a=\epbar_{+}\ga^a\psi_{m+}, \quad
\da^\rr_\ep\psi_{m+}=2 D^\rr_m(\wh\om^{+})\ep_{+}
\quad}
\eea
This shows that $(e_m{}^a, \psi_{m+})$ is the (standard)
2D $N=(1,0)$ supergravity multiplet.

\subsection{Radion multiplet}

In order to identify further submultiplets, we recall the basics
of the 2d $N=(1,0)$ supergravity tensor calculus \cite{ue2}. 
Besides the
supergravity multiplet we have just identified, there are two other
basic multiplets, the scalar multiplet $\Phi_2(A)=(A,\zeta_{-})$
and the spinor multiplet $\Psi_2(\zeta_{+})=(\zeta_{+},F)$.
They transform by definition as follows,
\bea
&& \da^\rr_\ep A=\epbar_{+}\zeta_{-}, \quad
\da^\rr_\ep\zeta_{-}=\ga^a\ep_{+}\wh D_a^\rr A \nn\\
&& \da^\rr_\ep \zeta_{+}=F\ep_{+}, \quad
\da^\rr_\ep F=\epbar_{+}\ga^a\wh D_a^\rr\zeta_{+}
\eea
where $\wh D_a^\rr A=\p_a A-\half\psibar_{a+}\chi_{-}$
and $\wh D_a^\rr\zeta_{+}=D_a^\rr(\wh\om^{+})\zeta_{+}
-\half F\psi_{a+}$ are supercovariant derivatives.

With these definitions, we now claim that
\bea
\label{rad}
\boxed{ \rule[-5pt]{0pt}{16pt} \quad
\Phi_2(e_3{}^{\hat 3})=(e_3{}^{\hat 3}, \quad 
-e_3{}^{\hat 3}\psi_{\hat 3-})
\quad}
\eea
is a good 2D $N=(1,0)$ scalar multiplet which we will call
the radion multiplet.\footnote{
The term ``radion'' refers to a field parametrizing the radius
of the extra dimension \cite{radion}. In our case, proper distances
in the $x^3$ direction must be measured with 
$g_{33}=e_3{}^{\hat 3} e_{3\hat 3}+e_3{}^a e_{3a}$, which is
not given by $e_3{}^{\hat 3}$ alone.
Nonetheless, we will call $\Phi(e_3{}^{\hat 3})$ the radion multiplet.
}
First of all, we observe that
$e_3{}^{\hat 3}$ is indeed a \emph{scalar} under the $\xi^m$ and $\la^{ab}$
transformations. The non-trivial part in this statement is that in
\bea
\da_\xi e_3{}^{\hat 3}=\xi^n\p_n e_3{}^{\hat 3} 
+e_n{}^{\hat 3}\p_3\xi^n
\eea
the last term vanishes in our gauge. Next, we apply the (modified)
$\ep_{+}$ susy to $e_3{}^{\hat 3}$ and find
\bea
\da^\rr_\ep e_3{}^{\hat 3}=\epbar_{+}\ga^{\hat 3}\psi_3
+\la^\rr{}^{\hat 3 a}e_{3a}
=\epbar_{+}(-\psi_{3-}+e_3{}^a\psi_{a-})
=\epbar_{+}(-e_3{}^{\hat 3}\psi_{\hat 3-})
\eea
which identifies the superpartner of $e_3{}^{\hat 3}$ as
$\zeta_{-}=-e_3{}^{\hat 3}\psi_{\hat 3-}$. To check that the
variation of $\zeta_{-}$ has the correct form is a bit more
involved. The details will be presented in \cite{tap}. 
The key intermediate statement is
\bea
\da^\rr_\ep\psi_{\hat 3-}
=P_{-}\Big[e_{\hat 3}{}^M\da\psi_M+\psi_M\da e_{\hat 3}{}^M \Big]
=\ga^a\ep_{+}
\Big[\wh\om_{\hat 3 a\hat 3}
-\half\psibar_{\hat 3+}\psi_{a-}\Big]
\eea
Next, in our gauge, it is easy to prove that
\bea
\wh\om_{\hat 3 a\hat 3}=-e_{\hat 3}{}^3\p_a e_{3\hat 3}
+\half(\psibar_{\hat 3+}\psi_{a-}-\psibar_{\hat 3-}\psi_{a+})
\eea
Finally, the contribution $\psi_{\hat 3-}\da e_3{}^{\hat 3}$ vanishes
thanks to the identity $(\epbar_{+}\psi_{-})\psi_{-}=0$.
Collecting the pieces, we find that $\da\zeta_{-}$ has the required form,
which proves that (\ref{rad}) is a good 2D $N=(1,0)$ scalar multiplet.

\subsection{Extrinsic curvature multiplet}

So far, we have found two \emph{even} submultiplets, the induced
supergravity and the radion multiplets. Now we will present an important
\emph{odd} submultiplet, the extrinsic curvature (scalar) multiplet.
The starting point is the (modified) $\ep_{+}$ susy transformation
of $\psi_{m-}$,
\bea
\da^\rr_\ep\psi_{m-}=\wh\om_{ma\hat 3}\ga^{a\hat 3}\ep_{+}
+\half\ga_m\ep_{+}S+\half\la^\rr_{a\hat 3}\ga^{a\hat 3}\psi_{m+}
\eea
Observing that $\da^\rr_\ep e_a{}^m=-(\epbar_{+}\ga^b\psi_{a+})e_b{}^m$,
we find, after some Fierzing, 
\bea
\da^\rr_\ep\psi_{a-}=\ga^b\ep_{+}\Big[
\wh K_{ab}+\half\eta_{ab}S \Big]
\eea
where $\wh K_{ab}$
is the (symmetric) supercovariant extrinsic curvature tensor
defined in (\ref{defK}). Contracting this expression with
$\ga^a$, we find
\bea
\da^\rr_\ep(\ga^a\psi_{a-})=(\wh K+S)\ep_{+}
\eea
where $\wh K=\eta^{ab}\wh K_{ab}$ is the (supercovariant)
extrinsic curvature scalar.
Noting that $\ga^a\psi_{a-}$ behaves as $\zeta_{+}$, we claim that
\bea
\label{ECM}
\boxed{ \rule[-5pt]{0pt}{16pt} \quad
\Psi_2(\ga^a\psi_{a-})=(\ga^a\psi_{a-}, \quad \wh K+S)
\quad}
\eea
is a good 2D $N=(1,0)$ spinor multiplet. The proof 
consists in demonstrating that
\bea
\da^\rr_\ep(\wh K+S)=\epbar_{+}\ga^a D_a^\rr(\wh\om^{+})[\ga^b\psi_{b-}]
-\half(\wh K+S)(\epbar_{+}\ga^a\psi_{a+})
\eea
The details of the proof will be presented in \cite{tap}, 
where we will also
discuss an extrinsic curvature \emph{tensor} multiplet
as well as a submultiplet that starts with $e_3{}^a$.

\subsection{Submultiplets of the 3D scalar multiplet}

In 3D $N=1$ supergravity, there is only one type of matter
multiplet, the scalar multiplet $\Phi_3(A)=(A,\chi,F)$.
(Other multiplets can be constructed by adding extra Lorentz
indices.) The susy transformations of this multiplet are
\bea
\label{susy2}
\da_\ep A=\epbar\chi, \quad
\da_\ep\chi=\ga^M\ep\wh D_M A+F\ep, \quad
\da_\ep F=\epbar\ga^M\wh D_M\chi-\frac{1}{4}S\epbar\chi
\eea
where $\wh D_M A=\p_M A-\half\psibar_M\chi$ and
$\wh D_M\chi=D_M(\wh\om)\chi-\half\ga^N\psi_M\wh D_N A
-\half F\psi_M$ are supercovariant derivatives.
Under the (modified) $\ep_{+}$ susy, this 3D multiplet splits
into the following two 2D $N=(1,0)$ submultiplets,\footnote{
We note that our co-dimension one multiplets contain terms of
the type ``odd $\cdot$ odd'' that are set to zero in the
approach of \cite{zucker,kugo}. For example, let us take
$F$ to be even, so that $\chi_{+}$ is even and $\chi_{-}$
is odd. The multiplet $\Psi_2(\chi_{+})$ is then even and contains
an explicit product of odd fields, $\psibar_{a-}\ga^a\chi_{-}$.
Such a product is also present in the radion multiplet (\ref{rad})
via the term $e_3{}^a\psi_{a-}$ inside 
$\zeta_{-}=-e_3{}^{\hat 3}\psi_{\hat 3-}$. 
For dimensions higher than 3D, such products also appear in the 
induced supergravity multiplet \cite{tap}.
}
\bea
\label{matter}
\boxed{ \rule[-5pt]{0pt}{16pt} \quad
\Phi_2(A)=(A, \;\; \chi_{-}), \qquad
\Psi_2(\chi_{+})=(\chi_{+}, \;\;
F+\wh D_{\hat 3}A-\half\psibar_{a-}\ga^a\chi_{-})
\quad}
\eea
The proof consists in showing that
\bea
\da^\rr_\ep A &=& \epbar_{+}\chi_{-}, \quad
\da^\rr_\ep\chi_{-} = \ga^a \ep_{+}\wh D_a^\rr A 
\nn\\[5pt]
\da^\rr_\ep\chi_{+} &=& F_2\ep_{+}, \quad
F_2\equiv F+\wh D_{\hat 3}A-\half\psibar_{a-}\ga^a\chi_{-}
\nn\\[5pt]
\da^\rr_\ep F_2 &=& \epbar_{+}\ga^a D^\rr(\wh\om^{+})_a\chi_{+}
-\half(\epbar_{+}\ga^a\psi_{a+}) F_2
\eea
where $\wh D_a^\rr A=e_a{}^m(\p_m A-\half\psibar_{m+}\chi_{-})$
and $\wh D_{\hat 3} A=e_{\hat 3}{}^M(\p_M A-\half\psibar_M\chi)$.
The proof is straightforward, except for the $\da^\rr_\ep F_2$ part
that we will discuss in \cite{tap}.

\subsection{Separately susy boundary actions}

In the 2D $N=(1,0)$ supergravity tensor calculus \cite{ue2}, 
susy actions are constructed
from spinor multiplets $\Psi_2(\zeta_{+})=(\zeta_{+},F)$
with the help of the following $F$-density formula,
\bea
\label{2DF}
\mc{L}_F\Big[\Psi_2(\zeta_{+})\Big]
=e_2\Big[F+\half\psibar_{a+}\ga^a\zeta_{+}\Big]
\eea
where $e_2=\det e_m{}^a$. In our case, this formula can be
directly applied to constructing (separately) susy invariant boundary
actions. Indeed, under the (modified) $\ep_{+}$ susy, we have
\bea
\da^\rr_\ep \mc{L}_F\Big[\Psi_2(\zeta_{+})\Big]
=\p_m\Big[ e_2(\epbar_{+}\ga^a\zeta_{+})e_a{}^m \Big]
\eea
and the total $\p_m$ derivative integrates to zero on the boundary.
Therefore,
\bea
\int_{\p\mc{M}} d^2x e_2\Big[F+\half\psibar_{a+}\ga^a\zeta_{+}\Big]
\eea
is a (separately) susy boundary action for a general spinor
multiplet $\Psi_2(\zeta_{+})=(\zeta_{+}, F)$. For example, we
can apply this formula to the extrinsic curvature multiplet
(\ref{ECM}) to obtain
\bea
\label{KSac}
\int_{\p\mc{M}} d^2x e_2\Big[
\wh K+S+\half\psibar_{a+}\ga^a\ga^b\psi_{b-} \Big]
\eea
which is (separately) supersymmetric under the (modified) 
$\ep_{+}$ susy (\ref{msusy}).

\section{Susy bulk-plus-boundary actions}

In this section, we will find an extension of the 3D $F$-density
formula that makes it very easy to construct susy bulk-plus-boundary
actions. We will then show how this formula can be written in terms of
co-dimension one submultiplets. Finally, we will use it 
to supersymmetrize the York-Gibbons-Hawking construction.

\subsection{The ``$F+A$'' formula}

In the 3D $N=1$ supergravity tensor calculus \cite{ue3}, 
susy actions are constructed
from scalar multiplets $\Phi_3(A)=(A,\chi,F)$
using the following $F$-density formula,
\bea
\mc{L}_F\Big[\Phi_3(A)\Big]=e_3\Big[
F+\half\psibar_M\ga^M\chi+\frac{1}{4}A\psibar_M\ga^{MN}\psi_N
+A S\Big]
\eea
where $e_3=\det e_M{}^A$.
Under 3D susy, this density transforms into a total 3D derivative,
\bea
\da_\ep \mc{L}_F\Big[\Phi_3(A)\Big]=\p_M\Big[
e_3\Big(\epbar\ga^M\chi+A\epbar\ga^{MN}\psi_N\Big) \Big]
\eea
In the presence of a boundary, 
the bulk $F$-density does not give rise
to a separately susy bulk action because the total derivative
yields a boundary term,
\bea
\int_\mc{M} d^3x \da_\ep \mc{L}_F\Big[\Phi_3(A)\Big]
=-\int_{\p\mc{M}} d^2x e_2 \Big(
\epbar\ga^{\hat 3}\chi+A\epbar\ga^{\hat 3 a}\psi_a \Big)
\eea
We used that, in our gauge, $e_a{}^3=0$ and 
$e_3 e_{\hat 3}{}^3=e_2$. Noting that $\mc{L}_F\Big[\Phi_3(A)\Big]$ 
is a Lorentz scalar, the (modified) $\ep_{+}$ susy transformation
(\ref{msusy}) gives
\bea
\label{remn}
\int_\mc{M} d^3x \da^\rr_\ep \mc{L}_F\Big[\Phi_3(A)\Big]
=\int_{\p\mc{M}} d^2x e_2 \Big(
\epbar_{+}\chi_{-}+A\epbar_{+}\ga^a\psi_{a+} \Big)
\eea
Noting that $\da^\rr_\ep A=\epbar_{+}\chi_{-}$ and
$\da^\rr_\ep e_2=e_2(\epbar_{+}\ga^a\psi_{a+})$, we can
construct a boundary action whose variation cancels (\ref{remn}).
The following bulk-plus-boundary action,
\bea
\label{F+A}
\boxed{ \rule[-5pt]{0pt}{16pt} \quad
S_{F+A} =
\int_\mc{M} d^3x \mc{L}_F\Big[\Phi_3(A)\Big]
-\int_{\p\mc{M}} d^2x e_2 A
\quad}
\eea
is invariant under the (modified) $\ep_{+}$ susy.
We call this the ``$F+A$'' formula.\footnote{
\label{nogauge}
The ``$F+A$'' formula (\ref{F+A}) has a natural extension 
to the case when the Lorentz gauge (\ref{gauge}) is not imposed 
\cite{tap}. We only have to replace $e_2=\det(e_m{}^a)$ 
with the determinant of the induced vielbein $e^\rr_2=\det(e^\rr_m{}^a)$
which satisfies 
$e^\rr_m{}^a e^\rr_{na}=e_m{}^a e_{na}+e_m{}^{\hat 3} e_{n\hat 3}$.
The resulting bulk-plus-boundary action is susy under the half of
bulk susy defined by $\ga^3\ep_{+}=\sqrt{g^{33}}\ep_{+}$. Note that
this makes the susy parameter $\ep_{+}$ \emph{field-dependent}
which makes the analysis of the gauge algebra more subtle
\cite{tap}. 
}

\subsection{Extended $F$-density}

As we will demonstrate explicitly in \cite{tap},
the boundary $A$-term can also be written as a bulk contribution
thanks to the following relation,
\bea
-\int_{\p\mc{M}} d^2x e_2 A
=\int_\mc{M} d^3x e_3(\p_{\hat 3} A+K A)
\eea
This allows us to define an extended $F$-density
\bea
\boxed{ \rule[-5pt]{0pt}{16pt} \quad
\mc{L}_F^\rr[\Phi_3(A)]=\mc{L}_F[\Phi_3(A)]+e_3(\p_{\hat 3} A+K A)
\quad}
\eea
whose integral over the bulk $\mc{M}$ reproduces the 
bulk-plus-boundary ``$F+A$'' formula (\ref{F+A}).
Under the (modified) $\ep_{+}$ susy, this extended 3D $F$-density 
behaves like the ordinary 2D $F$-density (that is, it varies into
a total $\p_m$ derivative). Therefore, we expect that it should be 
possible to rewrite it as a 2D $F$-density of some 2D $N=(1,0)$ 
spinor multiplet,\footnote{
In the superfield language, this corresponds to giving a prescription
for writing 3D \emph{locally} susy actions in terms of 2D superfields.
For \emph{rigid} susy, similar constructions are known in various 
dimensions \cite{rigidDd}. For the \emph{linearized} 5D supergravity,
the description in terms of 4D superfields was given in \cite{llp}.
For the full \emph{non-linear} 5D supergravity, such a construction
would require \cite{CST,db4} going beyond the orbifold supergravity 
tensor calculus of  \cite{zucker,kugo} where 
odd supergravity submultiplets
(like our extrinsic curvature multiplet (\ref{ECM})) and
``odd$\cdot$odd'' terms in even multiplets are discarded.
}
\bea
\mc{L}_F^\rr[\Phi_3(A)]=\mc{L}_F[\Psi_2(\zeta_{+})]
\eea
This is indeed possible, and we find \cite{tap}
\bea
\boxed{ \rule[-5pt]{0pt}{16pt} \quad
\Psi_2(\zeta_{+})=\Phi_2(e_3{}^{\hat 3})\times\Big[
\Psi_2(\chi_{+})+\Psi_2(\ga^a\psi_{a-})\times\Phi_2(A) \Big]
\quad}
\eea
where $\Phi_2(A)$ and $\Psi_2(\chi_{+})$ are the submultiplets 
(\ref{matter}) of the 3D scalar multiplet $\Phi_3(A)$, whereas
$\Phi_2(e_3{}^{\hat 3})$ and $\Psi_2(\ga^a\psi_{a-})$ are
the radion and the extrinsic curvature multiplets, respectively.
To derive this result, one needs the 
multiplication formula
\bea
(A,\; \zeta_{-})\times (\zeta_{+}, \;F)=
(A\zeta_{+}, \quad A F-\ov\zeta_{-}\zeta_{+})
\eea
which is part of the 2D $N=(1,0)$ tensor calculus \cite{ue2}.

\subsection{Super-York-Gibbons-Hawking construction}
\label{sec-ygh}

The ``$F+A$'' formula (\ref{F+A}) can be applied,
in particular, to the 3D scalar curvature multiplet,\footnote{
In our conventions, $R(\wh\om)=e_B{}^M e_A{}^N R(\wh\om)_{MN}{}^{AB}$
with $R(\wh\om)_{MN}{}^{AB}=\p_M\wh\om_N{}^{AB}
+\wh\om_M{}^{AC}\wh\om_{NC}{}^B-(M\leftrightarrow N)$,
and $\psi_{MN}=D_M(\wh\om)\psi_N-D_N(\wh\om)\psi_M$.
}
\bea
\Phi_3(S)= \Big( S, \quad
\frac{1}{2}\ga^{MN}\psi_{MN}-\half\ga^M\psi_M S, \quad
\frac{1}{2}R(\wh\om)-\frac{1}{2}\ov\psi^M\ga^N\psi_{MN}
+\frac{1}{4}S\ov\psi^M\psi_M-\frac{3}{4}S^2 \Big) \nn\\
\eea
We immediately obtain the following bulk-plus-boundary action,
\bea
\label{FASG}
S_{SG}=\int_\mc{M} d^3x e_3\Big[ \frac{1}{2}R(\wh\om)
+\frac{1}{2}\psibar_M\ga^{MNK}D(\wh\om)_N\psi_K
+\frac{1}{4}S^2 \Big]
-\int_{\p\mc{M}} d^2x e_2 S
\eea
which is, by construction, invariant under the (modified)
$\ep_{+}$ susy (without using any boundary conditions).
However, when one tries to apply the variational principle
to this action, one runs into a problem because the bulk auxiliary
field $S$ appears \emph{linearly} on the boundary.
(Its field equation would require $e_2$ to vanish, which is
too strong.)
This can be cured by adding a \emph{separately susy} boundary
action that removes the term linear in $S$. We add the action
given in (\ref{KSac}). The resulting improved bulk-plus-boundary
supergravity action reads\footnote{
The boundary term of the improved supergravity action (\ref{imprSG})
has the same form as the one found by Moss \cite{moss}.
(Note that
$2\psibar_{a+}\ga^a\ga^b\psi_{b-}=\psibar_a\ga^a\ga^b\psi_b$.)
However, there are essential differences.
Moss uses an ``adaptive coordinate system $e_{\hat N I}=\da_{N I}$,''
which in our case would mean $e_m{}^{\hat 3}=0$ \emph{and}
$e_3{}^{\hat 3}=1$. Moreover, his expression for the
supercovariant extrinsic curvature involves $\psi_N$ (our $\psi_3$)
and, therefore, could be equivalent to our (\ref{scEC}), 
which involves $\psi_{\hat 3}$, only if, in addition, $e_3{}^a=0$.
Finally, in the approach of Moss, susy of the bulk-plus-boundary
action is claimed only using the $\psi_{m-}=0$ boundary condition.
Our tensor calculus approach, on the other hand, leads to 
bulk-plus-boundary actions that are susy without using
any boundary conditions.
}
\bea
\label{imprSG}
S_{SG}^\text{impr} &=& \int_\mc{M} d^3x e_3\Big[ \frac{1}{2}R(\wh\om)
+\frac{1}{2}\psibar_M\ga^{MNK}D(\wh\om)_N\psi_K
+\frac{1}{4}S^2 \Big] \nn\\[5pt]
&& \quad+\int_{\p\mc{M}} d^2x e_2 \Big(
\wh K+\half\psibar_{a+}\ga^a\ga^b\psi_{b-} \Big)
\eea
where $\wh K=e^{ma}\wh K_{ma}$ with $\wh K_{ma}=\wh\om_{ma\hat 3}
-\half\psibar_{m+}\psi_{a-}$ which is the (symmetric) supercovariant
extrinsic curvature tensor.
The boundary term, which is obviously a susy generalization of
the York-Gibbons-Hawking term \cite{ygh}, 
can also be written as follows
\bea
\int_{\p\mc{M}} d^2x e_2 \Big(
\wt K+\half\psibar_{a+}\ga^{ab}\psi_{b-} \Big)
\eea
where $\wt K=e^{ma}\wt K_{ma}$ with $\wt K_{ma}=\wh\om_{ma\hat 3}$
which is neither symmetric nor supercovariant under the (modified)
$\ep_{+}$ susy.
The Euler-Lagrange variation of the improved 
supergravity action gives rise to the following boundary term,
\bea
\int_{\p\mc{M}} d^2x e_2 \Big[
\da e^{ma}(\wt K_{ma}-e_{ma}\wt K)
+\da\psibar_{m+}\ga^{ab}\psi_{b-} e_a{}^m \Big]
\eea
Therefore, removing the term linear in $S$ in the boundary action
of (\ref{FASG}) by adding a separately susy boundary action
(\ref{KSac}) has improved the variational principle it two ways. 
First, the unacceptable boundary condition $e_2=0$ is avoided. 
Second, the boundary part of the Euler-Lagrange variation
(known also as ``the boundary field equation'') is now in the
``$p\da q$'' form (by analogy with
the Hamiltonian formulation). 
This allows one to derive
``natural'' boundary conditions (for on-shell fields)
by requiring that the
boundary variation vanishes for arbitrary $\da q$ \cite{nbc}.
In our case, the role of ``$q$'' is played by the induced 
supergravity multiplet $(e_m{}^a, \psi_{m+})$ of (\ref{isg}).

It is very important for extending our construction to
higher dimensions (where the full set of auxiliary fields is
not always known or does not exist) 
that it is possible to eliminate the auxiliary
field $S$ by its equation of motion $S=0$ while preserving
susy of the action without the use of any boundary conditions.
This indicates, for example, that even though there is
no (off-shell) tensor calculus for 11D supergravity, the construction
of Moss \cite{moss} can, perhaps, be improved so that susy of the
11D Horava-Witten action on the manifold with boundary does not
require any boundary conditions on fields.

It is also instructive to find an alternative form of our
bulk-plus-boundary action (\ref{imprSG}) by separating 
the fermionic bilinear parts in $\wh\om_{MAB}$ and $\wh K$.
Setting $S=0$, we obtain \cite{tap}
\bea
\wt S_{SG} &=& 
\int_\mc{M} d^3x e_3\Big[ \frac{1}{2}R(\om)
+\frac{1}{2}\psibar_M\ga^{MNK}D(\om)_N\psi_K
+O(\psi^4) \Big] \nn\\[5pt]
&& \quad+\int_{\p\mc{M}} d^2x e_2 \Big(
K+\half\psibar_{a+}\ga^{ab}\psi_{b-} \Big)
\eea
where $K$ is the standard bosonic extrinsic curvature term.
In this form, ignoring the 4-fermi terms, the 3D bulk-plus-boundary
action for supergravity was
first found by Luckock and Moss in \cite{lmoss}.\footnote{
In 5D, the analog of this action was found in \cite{db2}
and its ``susy without BC'' was established up to the 4-fermi terms
and terms involving the 5D graviphoton. 
}
We have determined all 4-fermi terms in the bulk and boundary
actions. We found 4-fermi terms in the bulk action which agree
with the literature of supergravity, but no 4-fermi terms on
the boundary. So, the 2-fermi terms of \cite{lmoss} give
already the complete boundary action.
The new result of our construction is that the same boundary action
is sufficient for 
``susy without BC'' of the total bulk-plus-boundary action.

\section{Summary and Conclusions}

In this article, we have studied the issue of constructing
locally susy bulk-plus-boundary actions in the simple setting
of 3D $N=1$ supergravity. We demonstrated that the tensor
calculus for 3D $N=1$ supergravity can be naturally extended
to take boundaries into account. For a 3D scalar multiplet
$(A,\chi,F)$, our ``$F+A$'' formula (\ref{F+A}) gives a
bulk-plus-boundary action
\bea
S_{F+A}=\int_\mc{M} d^3x e_3\Big[ F+\dots\Big]
-\int_{\p\mc{M}} d^2x e_2 A
\eea
which is ``susy without BC'' (its susy variation vanishes
without the need to impose any BC on fields) under the half
of bulk susy parametrized by $\ep_{+}$ (satisfying 
$\ga^{\hat 3}\ep_{+}=\ep_{+}$ when the Lorentz gauge (\ref{gauge})
is imposed). Quite remarkably, this simple extension of
the standard $F$-density formula works in 4D $N=1$ 
sugra as well (where the $D$-density can also be
similarly extended) \cite{tap}. 

The ``$F+A$'' (extended $F$-density) formula can be applied
to a variety of models. As an illustration, we applied it
to the 3D $N=1$ scalar curvature multiplet. The resulting
bulk-plus-boundary action (\ref{FASG}) has the standard
3D $N=1$ sugra in the bulk and just the term $e_2 S$ on the
boundary. It is ``susy without BC'' by construction, but the
field equation for the bulk auxiliary field $S$ gives not only
$S=0$ in the bulk but also $e_2=0$ on the boundary, which is
unacceptable. To resolve this problem while maintaining 
the ``susy without BC'' property, we looked for an additional 
\emph{separately} susy boundary action containing the same term
$e_2 S$. The simplest such action is (\ref{KSac}).
Adding it to the minimal bulk-plus-boundary action given by
the ``$F+A$'' formula, we find that the $S$-term gets replaced
by the York-Gibbons-Hawking extrinsic curvature term $K$ together
with the gravitino bilinear $\psibar_{a+}\ga^{ab}\psi_{b-}$.
Neither the bulk nor the boundary action is separately susy,
but their sum is and it is ``susy without BC.''

In order to construct \emph{separately} susy boundary actions
systematically, we have developed a co-dimension one decomposition 
of bulk supermultiplets. We found that the 3D $N=1$ sugra multiplet
$(e_M{}^A,\psi_M,S)$ decomposes into several 2D $N=(1,0)$ multiplets:
the induced sugra multiplet $(e_m{}^a,\psi_{m+})$, the radion
multiplet $(e_3{}^{\hat 3}, -\psi_{3-}+e_3{}^a\psi_{a-})$ and
an ``off-diagonal multiplet'' 
$(e_{3a}, -e_3{}^{\hat 3}\psi_{a-}+\ga_a\psi_{3+})$ \cite{tap}.
(The other off-diagonal component of the vielbein, $e_m{}^{\hat 3}$, 
vanishes in our Lorentz gauge (\ref{gauge}).)
With the parity assignments given in (\ref{parity}),
the first two multiplets are ``even'' and the last one is ``odd.''
The 3D $N=1$ scalar multiplet $(A,\chi,F)$
allows a similar decomposition; see (\ref{matter}).
Explicit verification that these submultiplets transform
as standard 2D $N=(1,0)$ supermultiplets is tedious \cite{tap},
but our analysis of the gauge algebra guarantees that the 
co-dimension one decomposition does work and does not require
any (boundary) conditions on fields. 

In the superspace formulation, one can act on superfields with
superspace covariant derivatives to construct new superfields.
In the tensor calculus, the new multiplets can be constructed
simply by choosing an appropriate lowest component. For example,
starting with $\ga^a\psi_{a-}$, we obtain our extrinsic curvature
(scalar) multiplet (\ref{ECM}). Starting with $\psi_{a-}$, we
similarly obtain an extrinsic curvature \emph{tensor} multiplet
\cite{tap}. The multiplets obtained in this way can, together
with any number of independent boundary matter multiplets, be used
to construct separately susy boundary actions using the standard
2D $N=(1,0)$ $F$-density formula (\ref{2DF}). 
In conjunction with our ``$F+A$'' formula, this gives the
most general bulk-plus-boundary actions that are ``susy without BC.''
However, requiring that the variational principle yields field
equations that are not too strong restricts the choice of
boundary actions that one can allow \cite{tap}.

We should note that the Lorentz gauge (\ref{gauge}) that we
used in this work allows a tremendous simplification of the algebra.
At the same time, our results can be extended to the case when
no Lorentz gauge is imposed (see e.g. footnote~\ref{nogauge}) 
\cite{tap}. We also note that our tensor calculus approach
relies heavily on the \emph{off-shell} supergravity formulation
(with auxiliary fields). Such a formulation is not always available
in higher dimensions. Nonetheless, a concrete higher dimensional
model (such as the 11D Horava-Witten construction) has still a
chance to be ``susy without BC'' as we discussed in section 
\ref{sec-ygh}.

Our program of ``susy without BC'' can and should be extended to
(a) dimensions higher than three, (b) the superspace formulation,
(c) superconformal symmetries and superconformal actions, (d) BRST symmetry. 
Some progress in these directions has already
been achieved \cite{tap}. Ultimately, this would allow to have
complete control over the models discussed in the Introduction
as well as other models where symmetries and boundaries collide.

\vspace{30pt}
{\bf Acknowledgments.}
We would like to thank Dima Vassilevich for his participation
in the beginning of this project. D.V.B. also thanks Jon Bagger for
discussions on related topics. We thank the C. N. Yang Institute
for Theoretical Physics at SUNY Stony Brook and Deutsches
Electronen-Synchrotron DESY in Hamburg for hospitality extended
to us during visits related to this project. The research of D.V.B.
was supported in part by the German Science Foundation (DFG).
The research of P.v.N. was supported by the NSF grant no. PHY-0354776.




\begin{thebibliography}{999}


\bibitem{RTC}
  J.~Wess and B.~Zumino,
  Nucl.\ Phys.\  B {\bf 78}, 1 (1974).



\bibitem{LTC1}
  S.~Ferrara and P.~van Nieuwenhuizen,
  Phys.\ Lett.\  B {\bf 76}, 404 (1978);
  Phys.\ Lett.\  B {\bf 78}, 573 (1978).

\bibitem{LTC2}
  K.~S.~Stelle and P.~C.~West,
  Phys.\ Lett.\  B {\bf 77}, 376 (1978);
  Nucl.\ Phys.\  B {\bf 145}, 175 (1978).

\bibitem{SS}
  A.~Salam and J.~A.~Strathdee,
  Nucl.\ Phys.\  B {\bf 76}, 477 (1974);
  Phys.\ Rev.\  D {\bf 11}, 1521 (1975).

\bibitem{SSsg}
  J.~Wess and B.~Zumino,
  Phys.\ Lett.\  B {\bf 66}, 361 (1977).

\bibitem{LTCss}
  J.~Wess and B.~Zumino,
  Phys.\ Lett.\  B {\bf 79}, 394 (1978).


\bibitem{pdv}
  P.~Di Vecchia, B.~Durhuus, P.~Olesen and J.~L.~Petersen,
  Nucl.\ Phys.\  B {\bf 207}, 77 (1982);
%
  Nucl.\ Phys.\  B {\bf 217}, 395 (1983).

\bibitem{igarashi}
  Y.~Igarashi,
  Phys.\ Rev.\  D {\bf 30}, 1812 (1984);
%
  Y.~Igarashi and T.~Nonoyama,
  Phys.\ Lett.\  B {\bf 161}, 103 (1985);
%
  Phys.\ Rev.\  D {\bf 34} (1986) 1928.


\bibitem{bmods}
  S.~W.~Hawking,
  Phys.\ Lett.\  B {\bf 126}, 175 (1983);
%
  A.~K.~Chatterjee and P.~Majumdar,
  Phys.\ Lett.\  B {\bf 159}, 37 (1985);
%
  P.~D.~D'Eath,
  Nucl.\ Phys.\  B {\bf 269}, 665 (1986);
%
  S.~Elitzur, G.~W.~Moore, A.~Schwimmer and N.~Seiberg,
  Nucl.\ Phys.\  B {\bf 326}, 108 (1989);
%
  N.~Sakai and Y.~Tanii,
  Prog.\ Theor.\ Phys.\  {\bf 83}, 968 (1990);
%
  H.~Luckock,
  Annals Phys.\  {\bf 194}, 113 (1989);
%
  J.\ Phys.\ A  {\bf 24}, L1057 (1991);
%
  N.~P.~Warner,
  Nucl.\ Phys.\  B {\bf 450}, 663 (1995);
%
  T.~Inami, S.~Odake and Y.~Z.~Zhang,
  Phys.\ Lett.\  B {\bf 359}, 118 (1995);
%
  G.~Esposito and A.~Y.~Kamenshchik,
  Phys.\ Rev.\  D {\bf 54}, 3869 (1996);
%
  C.~Albertsson, U.~Lindstrom and M.~Zabzine,
  Commun.\ Math.\ Phys.\  {\bf 233}, 403 (2003);
%
  Nucl.\ Phys.\  B {\bf 678}, 295 (2004);
%
  U.~Lindstrom and M.~Zabzine,
  Phys.\ Lett.\  B {\bf 560}, 108 (2003);
%
  P.~Koerber, S.~Nevens and A.~Sevrin,
  JHEP {\bf 0311}, 066 (2003);
%
  S.~F.~Hassan,
  arXiv:hep-th/0308201;
%
  I.~V.~Melnikov, M.~R.~Plesser and S.~Rinke,
  arXiv:hep-th/0309223.




\bibitem{LRN}
  U.~Lindstrom, M.~Rocek and P.~van Nieuwenhuizen,
  Nucl.\ Phys.\  B {\bf 662}, 147 (2003).

\bibitem{VV}
  P.~van Nieuwenhuizen and D.~V.~Vassilevich,
  Class.\ Quant.\ Grav.\  {\bf 22}, 5029 (2005).


\bibitem{tap}
D.~V.~Belyaev and P.~van~Nieuwenhuizen, to appear.


\bibitem{db1}
  D.~V.~Belyaev,
  JHEP {\bf 0601}, 046 (2006).


\bibitem{rigidDd}
  N.~Marcus, A.~Sagnotti and W.~Siegel,
  Nucl.\ Phys.\  B {\bf 224}, 159 (1983);
%
  E.~A.~Mirabelli and M.~E.~Peskin,
  Phys.\ Rev.\ D {\bf 58}, 065002 (1998);
%
  N.~Arkani-Hamed, T.~Gregoire and J.~G.~Wacker,
  JHEP {\bf 0203}, 055 (2002);
%
  Y.~Sakamura,
  Nucl.\ Phys.\ B {\bf 656}, 132 (2003).


\bibitem{hw}
  P.~Horava and E.~Witten,
  Nucl.\ Phys.\  B {\bf 460}, 506 (1996);
%
  Nucl.\ Phys.\ B {\bf 475}, 94 (1996).


\bibitem{rs}
  L.~Randall and R.~Sundrum,
  Phys.\ Rev.\ Lett.\  {\bf 83}, 3370 (1999);
%
  Phys.\ Rev.\ Lett.\  {\bf 83}, 4690 (1999).


\bibitem{susyRS}
R.~Altendorfer, J.~Bagger and D.~Nemeschansky,
Phys.\ Rev.\ D {\bf 63}, 125025 (2001);
%
T.~Gherghetta and A.~Pomarol,
Nucl.\ Phys.\ B {\bf 586}, 141 (2000);
%
A.~Falkowski, Z.~Lalak and S.~Pokorski,
Phys.\ Lett.\ B {\bf 491}, 172 (2000);
%
  E.~Bergshoeff, R.~Kallosh and A.~Van Proeyen,
  JHEP {\bf 0010}, 033 (2000).


\bibitem{moss}
  I.~G.~Moss,
  Phys.\ Lett.\ B {\bf 577}, 71 (2003);
%
  Nucl.\ Phys.\  B {\bf 729}, 179 (2005).

\bibitem{db2}
  D.~V.~Belyaev,
  JHEP {\bf 0601}, 047 (2006).


\bibitem{zucker}
  M.~Zucker,
  Phys.\ Rev.\ D {\bf 64}, 024024 (2001);
%
  Fortsch.\ Phys.\  {\bf 51}, 899 (2003).

\bibitem{kugo}
  T.~Kugo and K.~Ohashi,
  Prog.\ Theor.\ Phys.\  {\bf 108}, 203 (2002).

\bibitem{db4}
  D.~V.~Belyaev,
  arXiv:0710.4540 [hep-th].



\bibitem{bbf}
  J.~Bagger and D.~V.~Belyaev,
  Phys.\ Rev.\ D {\bf 67}, 025004 (2003);
%
  JHEP {\bf 0306}, 013 (2003);
%
  Phys.\ Rev.\  D {\bf 72}, 065007 (2005);
%
  A.~Falkowski,
  JHEP {\bf 0505}, 073 (2005).



\bibitem{ue3}
  T.~Uematsu,
  Z.\ Phys.\ C {\bf 29}, 143 (1985);
%
  Z.\ Phys.\ C {\bf 32}, 33 (1986).


\bibitem{ue2}
  T.~Uematsu,
  Phys.\ Lett.\ B {\bf 183}, 304 (1987).



\bibitem{ygh}
J.~W.~York, Jr.,
Phys.\ Rev.\ Lett.\  {\bf 28}, 1082 (1972);
%
  G.~W.~Gibbons and S.~W.~Hawking,
  Phys.\ Rev.\ D {\bf 15}, 2752 (1977).


\bibitem{FPVN}
  D.~Z.~Freedman and P.~van Nieuwenhuizen,
  Phys.\ Rev.\  D {\bf 14}, 912 (1976).


\bibitem{KK}
  J.~F.~Luciani,
  Nucl.\ Phys.\  B {\bf 135}, 111 (1978);
%
  A.~Salam and J.~A.~Strathdee,
  Annals Phys.\  {\bf 141}, 316 (1982).

\bibitem{time1}
J.~Schwinger, 
Phys.\ Rev.\  {\bf 130}, 1253 (1963).

\bibitem{time2}
T.~W.~B.~Kibble,
J.\ Math.\ Phys.\  {\bf 4}, 1433 (1963).

\bibitem{hsg1}
  S.~Deser, J.~H.~Kay and K.~S.~Stelle,
  Phys.\ Rev.\  D {\bf 16}, 2448 (1977).

\bibitem{isenberg}
J.~Isenberg and J.~M.~Nester, ``Canonical Gravity,'' 
in \emph{General relativity and gravitation}, ed.~A.~Held, 
Plenum Press, New York, 1980.



\bibitem{radion}
  C.~Csaki, M.~Graesser, L.~Randall and J.~Terning,
  Phys.\ Rev.\  D {\bf 62}, 045015 (2000).


\bibitem{llp}
  W.~D.~Linch~III, M.~A.~Luty and J.~Phillips,
  Phys.\ Rev.\ D {\bf 68}, 025008 (2003).

\bibitem{CST}
  F.~Paccetti Correia, M.~G.~Schmidt and Z.~Tavartkiladze,
  Nucl.\ Phys.\ B {\bf 709}, 141 (2005).


\bibitem{nbc}
R.~Courant,
Bull.\ Amer.\ Math.\ Soc.\ {\bf 49}, 1 (1943);
%
  N.~H.~Barth,
  Class.\ Quant.\ Grav.\  {\bf 2}, 497 (1985).

\bibitem{lmoss}
  H.~Luckock and I.~Moss,
  Class.\ Quant.\ Grav.\  {\bf 6}, 1993 (1989).

\end{thebibliography}
\end{document}